\documentclass[10pt]{article}
\usepackage{graphicx}
\usepackage{amsmath}
\usepackage{amssymb}
\usepackage{caption2}
\begin{document}
\bibliographystyle{prsty}
\begin{center}
{\large {\bf \sc{  Analysis of  the strong decays of the $P_c(4312)$ as a pentaquark molecular state  with QCD sum rules }}} \\[2mm]
Zhi-Gang Wang \footnote{E-mail: zgwang@aliyun.com;\,\,zgwang@ncepu.edu.cn.  }, Xu Wang     \\
 Department of Physics, North China Electric Power University, Baoding 071003, P. R. China
\end{center}

\begin{abstract}
In this article, we tentatively assign the $P_c(4312)$ to be the $\bar{D}\Sigma_c$ pentaquark molecular state with the spin-parity $J^P={\frac{1}{2}}^-$, and discuss the
 factorizable  and non-factorizable contributions in the two-point QCD sum rules for  the $\bar{D}\Sigma_c$  molecular state in details to prove
 the reliability of the single pole approximation in the hadronic spectral density. We study its two-body strong decays with the QCD sum rules, special attentions are paid to match the hadron side with the QCD side of the correlation  functions to obtain solid duality.  We obtain the  partial decay  widths $\Gamma\left(P_c(4312)\to \eta_c p\right)=0.255\,{\rm{MeV}}$ and $\Gamma\left(P_c(4312)\to J/\psi p\right)=9.296^{+19.542}_{-9.296}\,\,{\rm{MeV}}$,  which
are compatible with the experimental value of the total width, and support assigning the $P_c(4312)$ to be the $\bar{D}\Sigma_c$ pentaquark molecular state.
\end{abstract}



\section{Introduction}

In 2015,  the  LHCb collaboration   observed  two pentaquark candidates $P_c(4380)$ and $P_c(4450)$ in the $J/\psi p$  mass spectrum  in the $\Lambda_b^0\to J/\psi K^- p$ decays \cite{LHCb-4380}. Recently, the LHCb collaboration  observed a  new narrow pentaquark candidate $P_c(4312)$ in the $ J/\psi  p$ mass spectrum with the statistical significance of  $7.3\sigma$, and confirmed the old $P_c(4450)$ pentaquark structure, which consists  of two narrow overlapping peaks $P_c(4440)$ and $P_c(4457)$
  with  the statistical significance of  $5.4\sigma$ \cite{LHCb-Pc4312}.
   The  masses and widths are
\begin{flalign}
 &P_c(4312) : M = 4311.9\pm0.7^{+6.8}_{-0.6} \mbox{ MeV}\, , \, \Gamma = 9.8\pm2.7^{+ 3.7}_{- 4.5} \mbox{ MeV} \, , \nonumber \\
 & P_c(4440) : M = 4440.3\pm1.3^{+4.1}_{-4.7} \mbox{ MeV}\, , \, \Gamma = 20.6\pm4.9_{-10.1}^{+ 8.7} \mbox{ MeV} \, , \nonumber \\
 &P_c(4457) : M = 4457.3\pm0.6^{+4.1}_{-1.7} \mbox{ MeV} \, ,\, \Gamma = 6.4\pm2.0_{- 1.9}^{+ 5.7} \mbox{ MeV} \,   .
\end{flalign}
  The $P_c(4312)$ can be assigned to be   a $\bar{D}\Sigma_c$ pentaquark molecular state \cite{Pc4312-molecule,WangPc4450-molecule},
  a pentaquark state \cite{Pc4312-di-di-anti, Pc4312-pentaquark, Pc4312-Wang-QCDSR}, a hadrocharmonium pentaquark state \cite{Pc4312-hadrocharmonium}.

 The $P_c(4312)$ lies near the $\bar{D}\Sigma_c$ threshold, which leads to the molecule assignment naturally.  In Ref.\cite{WangPc4450-molecule}, we perform detailed studies of  the $\bar{D}\Sigma_c$, $\bar{D}\Sigma_c^*$, $\bar{D}^{*}\Sigma_c$  and   $\bar{D}^{*}\Sigma_c^*$  pentaquark molecular states
with the QCD sum rules by carrying out the operator product expansion   up to   the vacuum condensates of dimension $13$ in a consistent way.
  The prediction $M_P=4.32\pm 0.11\,\rm{GeV}$ for the $\bar{D}\Sigma_c$ molecular state supports assigning the $P_c(4312)$ to be the $\bar{D}\Sigma_c$ pentaquark molecular state with the spin-parity $J^P={\frac{1}{2}}^-$. On the other hand, our studies based on the QCD sum rules  indicate that the scalar-diquark-scalar-diquark-antiquark type pentaquark state with the spin-parity $J^P={\frac{1}{2}}^-$ has a mass $4.31\pm 0.11\,\rm{GeV}$, the axialvector-diquark-axialvector-diquark-antiquark type pentaquark state with the spin-parity  $J^P={\frac{1}{2}}^-$ has a mass $4.34\pm 0.14\,\rm{GeV}$, which support assigning the $P_c(4312)$ to be a diquark-diquark-antiquark type pentaquark state \cite{Pc4312-Wang-QCDSR,Wang-Penta-2015}.
  The $P_c(4312)$ may be a diquark-diquark-antiquark type pentaquark state, which has a strong coupling to the $\bar{D}\Sigma_c$ scattering states, the strong coupling induces  some $\bar{D}\Sigma_c$ components \cite{Di-Wang-penta-mole}. So we can reproduce the experimental value of the mass of the $P_c(4312)$ in both  scenarios of the pentaquark state and pentaquark molecular state.
In Ref.\cite{Wang-Y4660-decay},  we  choose  the $[sc]_P[\bar{s}\bar{c}]_A-[sc]_A[\bar{s}\bar{c}]_P$  type tetraquark current to study the  strong decays of the  $Y(4660)$ with the QCD sum rules based on solid quark-hadron duality.
In calculations, we observe that the  hadronic coupling constants $ |G_{Y\psi^\prime f_0}|\gg |G_{Y J/\psi f_0}|$, which is consistent with the observation of the $Y(4660)$ in the $\psi^\prime\pi^+\pi^-$ mass spectrum, and favors the $\psi^{\prime}f_0(980)$ molecule assignment \cite{Wang-CTP-4660}. The similar  mechanism maybe exist for the $P_c(4312)$.

  In this article, we tentatively assign the $P_c(4312)$ to be the $\bar{D}\Sigma_c$ pentaquark molecular state with the spin-parity  $J^P={\frac{1}{2}}^-$, and study its two-body strong decays with the QCD sum rules.  In Ref.\cite{WangZhang-Solid}, we assign  the $Z_c(3900)$ to be  the diquark-antidiquark  type axialvector  tetraquark  state,
 study the hadronic coupling  constants in the strong decays $Z_c(3900) \to J/\psi\pi$, $\eta_c\rho$, $D \bar{D}^{*}$ with the QCD sum rules based on solid quark-hadron duality  by taking  into account both the connected and disconnected Feynman diagrams in the operator product expansion.  The method  works well in studying the two-body strong decays of the $Z_c(3900)$, $X(4140)$, $X(4274)$ and $Z_c(4600)$  \cite{WangZhang-Solid,Wang-Y4140-Y4274}. Now we extend the method to study the two-body strong decays of the pentaquark molecular state by carrying out the operator product expansion up to the vacuum condensates of dimension $10$.

 The article is arranged as follows:  in Sect.2, we present  comments on the QCD sum rules for  the $\bar{D}\Sigma_c$  pentaquark molecular state; in Sect.3, we derive the QCD sum rules for the  hadronic  coupling constants in the strong decays $P_c(4312)\to \eta_c p$, $J/\psi p$;    in Sect.4, we present the numerical results and discussions; and Sect.5 is reserved for our
conclusion.

\section{Comments on the QCD sum rules for  the $\bar{D}\Sigma_c$  pentaquark molecular state}

In the following, we write down  the two-point correlation function $\Pi(p)$ to study the mass and pole residue of the $\bar{D}\Sigma_c$  pentaquark molecular state
with the QCD sum rules,
\begin{eqnarray}
\Pi(p)&=&i\int d^4x e^{ip \cdot x} \langle0|T\left\{J(x)\bar{J}(0)\right\}|0\rangle \, ,
\end{eqnarray}
where the current $J(x)=J_{\bar{D}\Sigma_c}(x)$,
\begin{eqnarray}
 J_{\bar{D}\Sigma_c}(x)&=& \bar{c}(x)i\gamma_5 u(x)\, \varepsilon^{ijk}  u^T_i(x) C\gamma_\alpha d_j(x)\, \gamma^\alpha\gamma_5 c_{k}(x) \, ,
\end{eqnarray}
the $i$, $j$, $k$ are color indices. We choose the color-singlet-color-singlet type (or meson-baryon type) current  $J_{\bar{D}\Sigma_c}(x)$ to interpolate the $\bar{D}\Sigma_c$ pentaquark molecular state with the spin-parity $J^P={\frac{1}{2}}^-$ \cite{WangPc4450-molecule}. For the technical details  and  numerical results, one can consult Ref.\cite{WangPc4450-molecule}. In the present work, we will focus on the reliability of the single pole approximation in the hadronic spectral density.

At the QCD side, the correlation function $\Pi(p)$ can be written as
\begin{eqnarray}\label{Pi12}
\Pi(p)&=&-i\, \varepsilon^{ijk} \varepsilon^{i^{\prime}j^{\prime}k^{\prime}}
 \int d^4x e^{ip\cdot x}\, \nonumber\\
&&\Big\{   -{\rm Tr}\left[i\gamma_5 C_{m^\prime m}(-x) i\gamma_5  U_{mm^\prime}(x)\right] \,{\rm Tr}\left[\gamma_\alpha D_{jj^\prime}(x) \gamma_\beta C U^{T}_{ii^\prime}(x)C\right] \gamma^\alpha \gamma_5 C_{kk^{\prime}}(x)\gamma_5\gamma^\beta \nonumber\\
&& +  {\rm Tr} \left[i\gamma_5 C_{m^\prime m}(-x) i\gamma_5  U_{mi^\prime}(x) \gamma_\beta C D^T_{jj^\prime}(x)C \gamma_\alpha  U_{im^\prime}(x)\right] \gamma^\alpha \gamma_5 C_{kk^{\prime}}(x)\gamma_5\gamma^\beta   \Big\} \, ,
\end{eqnarray}
where
the $U_{ij}(x)$, $D_{ij}(x)$ and $C_{ij}(x)$ are the full $u$, $d$ and $c$ quark propagators respectively ($S_{ij}(x)=U_{ij}(x),\,D_{ij}(x)$),
 \begin{eqnarray}\label{Lpropagator}
S_{ij}(x)&=& \frac{i\delta_{ij}\!\not\!{x}}{ 2\pi^2x^4}-\frac{\delta_{ij}\langle
\bar{q}q\rangle}{12} -\frac{\delta_{ij}x^2\langle \bar{q}g_s\sigma Gq\rangle}{192} -\frac{ig_sG^{a}_{\alpha\beta}t^a_{ij}(\!\not\!{x}
\sigma^{\alpha\beta}+\sigma^{\alpha\beta} \!\not\!{x})}{32\pi^2x^2} \nonumber\\
&&  -\frac{1}{8}\langle\bar{q}_j\sigma^{\mu\nu}q_i \rangle \sigma_{\mu\nu}+\cdots \, ,
\end{eqnarray}
\begin{eqnarray}\label{Hpropagator}
C_{ij}(x)&=&\frac{i}{(2\pi)^4}\int d^4k e^{-ik \cdot x} \left\{
\frac{\delta_{ij}}{\!\not\!{k}-m_c}
-\frac{g_sG^n_{\alpha\beta}t^n_{ij}}{4}\frac{\sigma^{\alpha\beta}(\!\not\!{k}+m_c)+(\!\not\!{k}+m_c)
\sigma^{\alpha\beta}}{(k^2-m_c^2)^2}\right.\nonumber\\
&&\left. -\frac{g_s^2 (t^at^b)_{ij} G^a_{\alpha\beta}G^b_{\mu\nu}(f^{\alpha\beta\mu\nu}+f^{\alpha\mu\beta\nu}+f^{\alpha\mu\nu\beta}) }{4(k^2-m_c^2)^5}+\cdots\right\} \, ,\nonumber\\
f^{\alpha\beta\mu\nu}&=&(\!\not\!{k}+m_c)\gamma^\alpha(\!\not\!{k}+m_c)\gamma^\beta(\!\not\!{k}+m_c)\gamma^\mu(\!\not\!{k}+m_c)\gamma^\nu(\!\not\!{k}+m_c)\, ,
\end{eqnarray}
and  $t^n=\frac{\lambda^n}{2}$, the $\lambda^n$ is the Gell-Mann matrix   \cite{PRT85,Pascual-1984,WangHuangTao-3900}.

\begin{figure}
 \centering
  \includegraphics[totalheight=4cm,width=10cm]{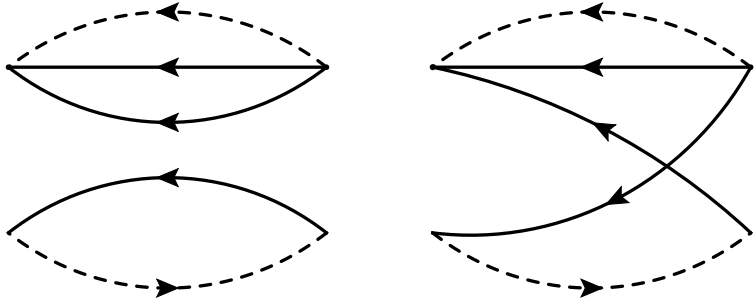}
 \caption{ The  Feynman diagrams  for the lowest order  contributions for the baryon-meson type current $J(x)$, where the solid lines and dashed lines denote the light quarks and heavy quarks, respectively. We have  taken into account the finite spatial  separation between the $\bar{c}(x)i\gamma_5 u(x)$ and $\varepsilon^{ijk}  u^T_i(x) C\gamma_\alpha d_j(x)\, \gamma^\alpha\gamma_5 c_{k}(x)$ clusters in the current operator $J(x)$.}\label{Lowest-diagram}
\end{figure}

In Fig.\ref{Lowest-diagram}, we plot the two Feynman diagrams  for  the lowest order  contributions, where the first diagram corresponds the term with two Tr's  and the
second diagram corresponds to the term  with one Tr in Eq.\eqref{Pi12}. The first Feynman diagram is factorizable and has the color factor $18$,
 while the second Feynman diagram is non-factorizable and  has the color factor $6$. In the large $N_c$ limit $N_c \to \infty$, the contribution of the second Feynman diagram is greatly suppressed. In reality, the color number $N_c=3$, the second Feynman diagram plays an important role.

In the second Feynman diagram, we can replace the lowest order heavy quark lines and (or) light quark lines with other terms in the full propagators in Eqs.\eqref{Lpropagator}-\eqref{Hpropagator}, and obtain other non-factorizable Feynman diagrams.

\begin{figure}
 \centering
  \includegraphics[totalheight=4cm,width=10cm]{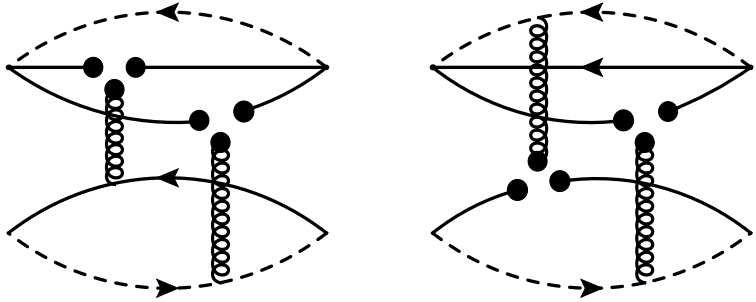}
 \caption{ The non-factorizable  Feynman diagrams contributing to the vacuum condensates
$\langle \bar{q}g_s\sigma G q \rangle^2$ from the terms with two Tr's in Eq.\eqref{Pi12}, where the solid lines and dashed lines denote the light quarks and heavy quarks, respectively. Other
diagrams obtained by interchanging of the  light quark lines  are implied.  }\label{diagram-qqg-qqg}
\end{figure}

\begin{figure}
 \centering
  \includegraphics[totalheight=4cm,width=10cm]{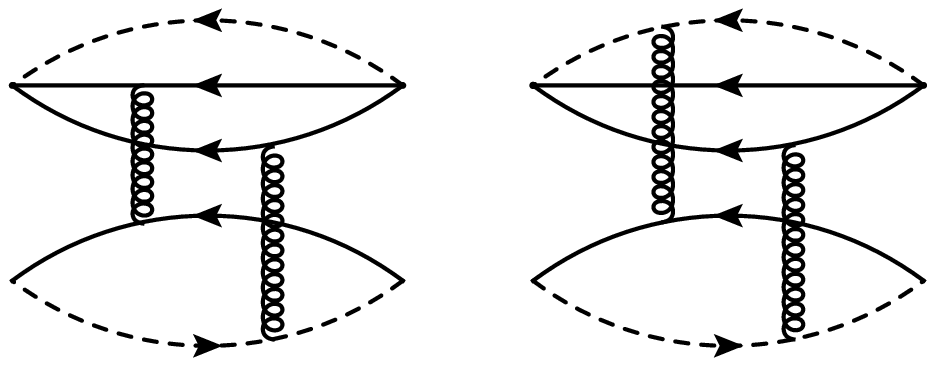}\\
    \vspace{1cm}
    \includegraphics[totalheight=4cm,width=10cm]{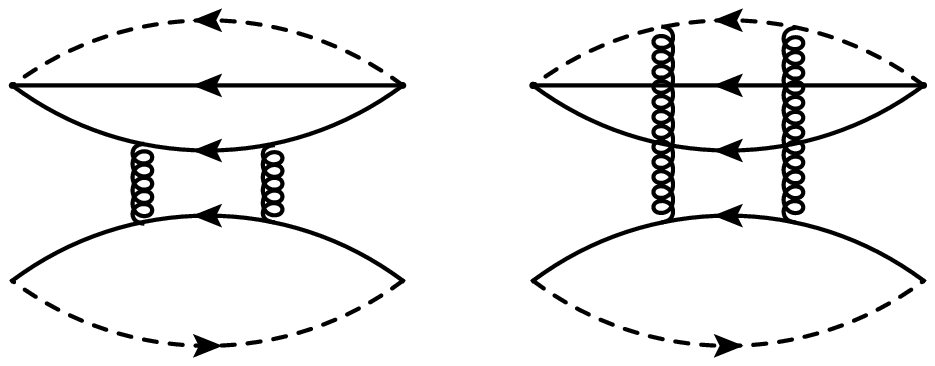}\\
    \vspace{1cm}
    \includegraphics[totalheight=4cm,width=10cm]{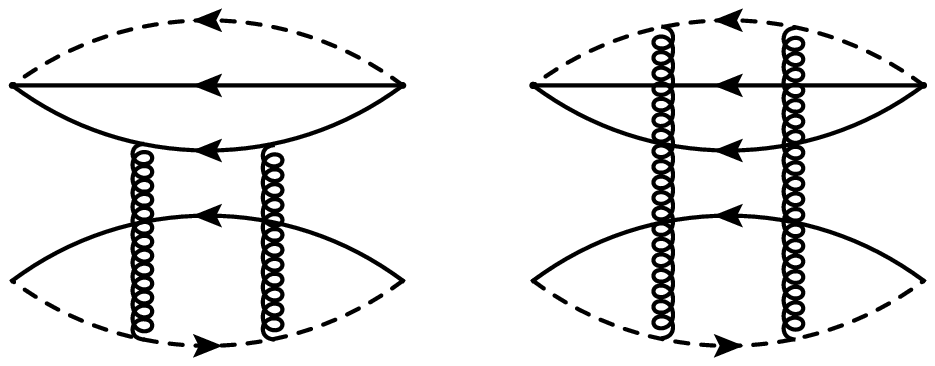}\\
    \vspace{1cm}
 \caption{ The non-factorizable  Feynman diagrams of the order $\mathcal{O}(\alpha_s^2)$ from the terms with two Tr's in Eq.\eqref{Pi12}, where the solid lines and dashed lines denote the light quarks and heavy quarks, respectively. Other
diagrams obtained by interchanging of the  light and heavy quark lines  are implied.  }\label{diagram-afs-afs}
\end{figure}

In the first Feynman diagram, we can also replace the lowest order heavy quark lines and (or) light quark lines with other terms in the full propagators in Eqs.\eqref{Lpropagator}-\eqref{Hpropagator}, and obtain other factorizable Feynman diagrams. There are non-factorizable Feynman diagrams besides the factorizable Feynman diagrams, see Fig.\ref{diagram-qqg-qqg}. In Fig.\ref{diagram-qqg-qqg}, we plot the Feynman diagrams contributing to the vacuum condensates
$\langle \bar{q}g_s\sigma G q \rangle^2$, which are the vacuum expectations of the quark-gluon operators of the order $\mathcal{O}(\alpha_s)$, not of the order $\mathcal{O}(\alpha_s^2)$. In Fig.\ref{diagram-afs-afs}, we plot the non-factorizable  Feynman diagrams of the order $\mathcal{O}(\alpha_s^2)$ from the terms with two Tr's in Eq.\eqref{Pi12},  the first, second, third and fourth diagrams are non-planar Feynman  diagrams, while  the fifth and sixth diagrams are planar Feynman  diagrams. The  first and second Feynman diagrams  are suppressed by a factor $\frac{1}{\sqrt{N_c}^4 }\frac{1}{N_c}=\frac{1}{N_c^3}$ in the large $N_c$ limit compared to the first Feynman  diagram in Fig.\ref{Lowest-diagram}, while the third, fourth, fifth and sixth diagrams are suppressed by a factor $\frac{1}{\sqrt{N_c}^4 }=\frac{1}{N_c^2}$. In reality, the color number $N_c=3$, the Feynman diagrams in Fig.\ref{diagram-afs-afs} are suppressed by a factor $(\frac{4}{3}\frac{\alpha_s}{4\pi})^2\sim0.0009$, and play a minor important  role.

\begin{figure}
 \centering
  \includegraphics[totalheight=4cm,width=5cm]{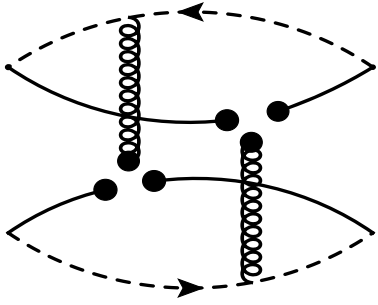}
 \caption{ The non-factorizable  Feynman diagrams contributing to the vacuum condensates
$\langle \bar{q}g_s\sigma G q \rangle^2$ for the meson-meson type currents, where the solid lines and dashed lines denote the light quarks and heavy quarks, respectively. }\label{meson-qqg-qqg}
\end{figure}

In Fig.\ref{meson-qqg-qqg}, we plot the non-factorizable  Feynman diagrams contributing to the vacuum condensates
$\langle \bar{q}g_s\sigma G q \rangle^2$ for the meson-meson type currents. From the figure, we can see that the non-factorizable contributions begin at the order $\mathcal{O}(\alpha_s^0)$ rather than at  the order $\mathcal{O}(\alpha_s^2)$ argued in Ref.\cite{Three-Chu-Sheng}. For the nonperturbative contributions, we absorb the strong coupling constant $g_s^2=4\pi\alpha_s$ into the vacuum condensates and count them as of the order $\mathcal{O}(\alpha_s^0)$.

We  insist on  the viewpoint that the  factorizable Feynman diagrams correspond  to the two-particle reducible contributions,
 irrespective of the baryon-meson pair or the meson-meson pair,
  and give the masses of the two constituent particles, then the attractive interactions which originate from (or are embodied in) the non-factorizable Feynman diagrams attract the two  constituent particles to form the molecular states.
 The non-factorizable Feynman diagrams are suppressed in the large $N_c$ limit, which is consistent with the small bound energies  of the  pentaquark molecular states.   The baryon-meson type or color-singlet-color-singlet type currents   couple potentially to the pentaquark molecular states.

On the other hand, the baryon-meson type currents  also couple  to the baryon-meson pairs besides the molecular states
as there exist two-particle reducible contributions,
the intermediate  baryon-meson loops  contribute   a finite imaginary part to modify the dispersion relation at the hadron side \cite{WangPc4450-molecule}. In calculations, we observe that   the zero width approximation works well, the couplings to the baryon-meson pairs  can be neglected safely.

  If we only take into account the non-factorizable Feynman diagrams shown Figs.\ref{Lowest-diagram}-\ref{diagram-qqg-qqg}, even if we obtain  stable QCD sum rules, we cannot distinguish the diquark-diquark-antiquark type substructure or the baryon-meson type substructure, and  cannot select the color-singlet-color-singlet type substructure and refer to it as the molecular state, we just obtain a hidden-charm  five-quark state with the spin-parity  $J^P={\frac{1}{2}}^-$.
  If we insist on that it is a molecular state, which diagram contributes to masses of the baryon and meson constituents?
  In Ref.\cite{Three-Chu-Sheng-Baryon},
 the factorizable Feynman diagrams corresponding   to the two-particle reducible contributions are subtracted, only the non-factorizable Feynman diagrams are taken into account to study the pentaquark states. We do not agree with that approach.

\section{QCD sum rules for  the $P_c(4312)$ decays as a  pentaquark molecular state}

In the following, we write down  the three-point correlation functions $\Pi_5(p,q)$ and $\Pi_{\mu}(p,q)$   in the QCD sum rules,
\begin{eqnarray}
\Pi_5(p,q)&=&i^2\int d^4xd^4y e^{ip \cdot x}e^{iq \cdot y} \langle0|T\left\{J_5(x)J_N(y)\bar{J}(0)\right\}|0\rangle \, , \\
\Pi_\mu(p,q)&=&i^2\int d^4xd^4y e^{ip \cdot x}e^{iq \cdot y} \langle0|T\left\{J_\mu(x)J_N(y)\bar{J}(0)\right\}|0\rangle \, ,
\end{eqnarray}
where
\begin{eqnarray}
 J_5(x)&=& \bar{c}(x)i\gamma_5 c(x)\, ,\nonumber\\
 J_\mu(x)&=& \bar{c}(x) \gamma_\mu c(x)\, ,\nonumber\\
 J_N(y)&=&  \varepsilon^{ijk}  u^T_i(y) C\gamma_\alpha u_j(y)\, \gamma^\alpha\gamma_5 d_{k}(y) \, ,\nonumber \\
 J(0)&=& \bar{c}(0)i\gamma_5 u(0)\, \varepsilon^{ijk}  u^T_i(0) C\gamma_\alpha d_j(0)\, \gamma^\alpha\gamma_5 c_{k}(0) \, ,
\end{eqnarray}
the $i$, $j$, $k$ are color indices. We choose the currents $J_5(x)$, $J_\mu(x)$, $J_N(y)$ and $ J(0)$ to interpolate the $\eta_c$, $J/\psi$, $p$ and $P_c(4312)$, respectively. Thereafter we will denote the proton $p$ as $N$ to avoid confusion due to the four momentum $p_\mu$.

 At the hadron  side, we  insert  a complete set  of intermediate hadron  states with the
same quantum numbers as the current operators $J_5(x)$, $J_\mu(x)$, $J_N(y)$ and $ J(0)$  into the correlation functions $\Pi_5(p,q)$ and $\Pi_{\mu}(p,q)$
 to obtain the hadronic representation \cite{PRT85,SVZ79}.  After isolating the pole terms of the ground  states, we obtain the
following results:
\begin{eqnarray}
 \Pi_5(p,q) & = & \frac{f_{\eta_c}m_{\eta_c}^2\lambda_P\lambda_N}{2m_c}\frac{-iu(q)\,\langle\eta_c(p)N(q)|P(p^\prime)\rangle \, \bar{u}(p^\prime)}{\left(m_P^2-p^{\prime2}\right)\left(m_{\eta_c}^2-p^{2}\right)\left(m_{N}^2-q^{2}\right)}+\cdots\, \nonumber\\
  & = & \frac{f_{\eta_c}m_{\eta_c}^2\lambda_P\lambda_N}{2m_c}\frac{-i \left(\!\not\!{q}+m_N \right)\left( \!\not\!{p}^\prime+m_P \right)}{\left(m_P^2-p^{\prime2}\right)\left(m_{\eta_c}^2-p^{2}\right)\left(m_{N}^2-q^{2}\right)}ig_5+\cdots\,  ,
\end{eqnarray}

\begin{eqnarray}
 \Pi_\mu(p,q) & = & f_{J/\psi}m_{J/\psi}\lambda_P\lambda_N\frac{-i \varepsilon_\mu u(q)\,\langle J/\psi(p)N(q)|P(p^\prime)\rangle \, \bar{u}(p^\prime)}{\left(m_P^2-p^{\prime2}\right)\left(m_{J/\psi}^2-p^{2}\right)\left(m_{N}^2-q^{2}\right)}+\cdots\, \nonumber\\
  & = & f_{J/\psi}m_{J/\psi}\lambda_P\lambda_N\frac{-i \left(\!\not\!{q}+m_N \right)\left( g_V\gamma^\alpha-i\frac{g_T}{m_P+m_N}\sigma^{\alpha\beta}p_\beta\right)\gamma_5\left( \!\not\!{p}^\prime+m_P \right)}{\left(m_P^2-p^{\prime2}\right)\left(m_{J/\psi}^2-p^{2}\right)\left(m_{N}^2-q^{2}\right)}\nonumber\\
  &&\left(-g_{\mu\alpha}+\frac{p_{\mu}p_{\alpha}}{p^2} \right)+\cdots\,  ,
\end{eqnarray}
where we have used the definitions,
\begin{eqnarray}
\langle 0| J (0)|P_{c}(p^\prime)\rangle &=&\lambda_{P} U(p^\prime,s) \, ,  \nonumber\\
\langle 0| J_{N} (0)|N(q)\rangle &=&\lambda_{N} U(q,s) \, , \nonumber \\
\langle 0| J_{\mu} (0)|J/\psi(p)\rangle &=&f_{J/\psi}m_{J/\psi}\varepsilon_\mu(p,s) \, , \nonumber \\
\langle 0| J_{5} (0)|\eta_c(p)\rangle &=&\frac{f_{\eta_c}m_{\eta_c}^2}{2m_c} \, ,
\end{eqnarray}
\begin{eqnarray}\label{Coupling-gVgT}
\langle\eta_c(p)N(q)|P(p^\prime)\rangle&=&i g_5\bar{u}(q)u(p^\prime)\, ,\nonumber\\
\langle J/\psi(p)N(q)|P(p^\prime)\rangle&=& \bar{u}(q)\varepsilon^*_\alpha\left( g_V\gamma^\alpha-i\frac{g_T}{m_P+m_N}\sigma^{\alpha\beta}p_\beta\right)\gamma_5u(p^\prime)\, ,
\end{eqnarray}
the $g_5$, $g_V$ and $g_T$ are the hadronic coupling constants, the $U(p,s)$ and $U(q,s)$ are the Dirac spinors, the $\lambda_P$ and $\lambda_N$ are the pole residues,
the $f_{J/\psi}$ and $f_{\eta_c}$ are the decay constants, the $\varepsilon_\mu$ is the polarization vector of the $J/\psi$.

It is important to choose the pertinent structures to study the hadronic coupling constants.
If $\Pi_{5,H}(p,q)=\Pi_{5,QCD}(p,q)$ and $\Pi_{\mu,H}(p,q)=\Pi_{\mu,QCD}(p,q)$, we expect that the two relations ${\rm Tr}\left[\Pi_{5,H}(p,q)\Gamma\right]={\rm Tr}\left[\Pi_{5,QCD}(p,q)\Gamma\right]$ and ${\rm Tr}\left[\Pi_{\mu,H}(p,q)\Gamma^\prime\right]={\rm Tr}\left[\Pi_{\mu,QCD}(p,q)\Gamma^\prime\right]$ also exist, where the subscripts $H$ and $QCD$ denote the  hadron side and   QCD side of the correlation functions, respectively, the $\Gamma$ and $\Gamma^\prime$ are some Dirac
$\gamma$-matrixes.

In this article, we choose $\Gamma=\sigma_{\mu\nu}$, $i\gamma_\mu$, $\Gamma^\prime=\gamma_5\!\not\!{z}$, $\gamma_5$,
\begin{eqnarray}
\frac{1}{4}{\rm Tr}\left[\Pi_5(p,q)\sigma_{\mu\nu}\right]&=& \Pi_5(p^{\prime2},p^2,q^2) \,i\left(p_\mu q_\nu-q_\mu p_\nu \right)+\cdots\, , \nonumber\\
\frac{1}{4}{\rm Tr}\left[\Pi_5(p,q)i\gamma_\mu  \right]&=& \overline{\Pi}_5(p^{\prime2},p^2,q^2)\, iq_\mu  +\cdots\, , \nonumber\\
\frac{1}{4}{\rm Tr}\left[\Pi_\mu(p,q)\gamma_5\!\not\!{z} \right]&=& \Pi_A(p^{\prime2},p^2,q^2)\, iq_\mu p\cdot z+\cdots\, ,\nonumber\\
\frac{1}{4}{\rm Tr}\left[\Pi_\mu(p,q)\gamma_5  \right]&=& \Pi_B(p^{\prime2},p^2,q^2)\, iq_\mu  +\cdots\, ,
\end{eqnarray}
 and choose the tensor structures $p_\mu q_\nu-q_\mu p_\nu$, $q_\mu$, $q_\mu p\cdot z$ and $q_\mu$ to study the hadronic coupling constants $g_5$, $g_V$ and $g_T$, respectively,
 where the $z_\mu$ is a four vector.

Now we write down the components $\Pi_5(p^{\prime2},p^2,q^2)$, $\overline{\Pi}_5(p^{\prime2},p^2,q^2)$, $\Pi_A(p^{\prime2},p^2,q^2)$ and $\Pi_B(p^{\prime2},p^2,q^2)$ explicitly,
\begin{eqnarray}\label{Hadron-Pi5}
 \Pi_5(p^{\prime2},p^2,q^2)   & = & \frac{f_{\eta_c}m_{\eta_c}^2\lambda_P\lambda_N}{2m_c}\frac{g_5}{\left(m_P^2-p^{\prime2}\right)\left(m_{\eta_c}^2-p^{2}\right)\left(m_{N}^2-q^{2}\right)}\nonumber\\
 &&+ \frac{1}{(m_{P}^2-p^{\prime2})(m_{\eta_c}^2-p^2)} \int_{s^0_N}^\infty dt\frac{\rho^5_{PN^\prime}(p^{\prime 2},p^2,t)}{t-q^2}\nonumber\\
&& + \frac{1}{(m_{P}^2-p^{\prime2})(m_{N}^2-q^2)} \int_{s^0_{\eta_c}}^\infty dt\frac{\rho^5_{P\eta_c^\prime}(p^{\prime 2},t,q^2)}{t-p^2}  \nonumber\\
&& + \frac{1}{(m_{\eta_c}^2-p^{2})(m_{N}^2-q^2)} \int_{s^0_{P}}^\infty dt\frac{\rho^5_{P^{\prime}\eta_c}(t,p^2,q^2)+\rho^5_{P^{\prime}N}(t,p^2,q^2)}{t-p^{\prime2}}+\cdots \, ,
\end{eqnarray}

\begin{eqnarray}
 \overline{\Pi}_5(p^{\prime2},p^2,q^2)   & = & \frac{f_{\eta_c}m_{\eta_c}^2\lambda_P\lambda_N}{2m_c}\frac{\left(m_P+m_N\right)g_5}{\left(m_P^2-p^{\prime2}\right)\left(m_{\eta_c}^2-p^{2}\right)\left(m_{N}^2-q^{2}\right)}\nonumber\\
 &&+ \frac{1}{(m_{P}^2-p^{\prime2})(m_{\eta_c}^2-p^2)} \int_{s^0_N}^\infty dt\frac{\overline{\rho}^5_{PN^\prime}(p^{\prime 2},p^2,t)}{t-q^2}\nonumber\\
&& + \frac{1}{(m_{P}^2-p^{\prime2})(m_{N}^2-q^2)} \int_{s^0_{\eta_c}}^\infty dt\frac{\overline{\rho}^5_{P\eta_c^\prime}(p^{\prime 2},t,q^2)}{t-p^2}  \nonumber\\
&& + \frac{1}{(m_{\eta_c}^2-p^{2})(m_{N}^2-q^2)} \int_{s^0_{P}}^\infty dt\frac{\overline{\rho}^5_{P^{\prime}\eta_c}(t,p^2,q^2)+\overline{\rho}^5_{P^{\prime}N}(t,p^2,q^2)}{t-p^{\prime2}}+\cdots \, ,
\end{eqnarray}

\begin{eqnarray}
 \Pi_A(p^{\prime2},p^2,q^2)   & = & f_{J/\psi}m_{J/\psi}\lambda_P\lambda_N\frac{g_T-g_V}{\left(m_P^2-p^{\prime2}\right)\left(m_{J/\psi}^2-p^{2}\right)\left(m_{N}^2-q^{2}\right)} \nonumber\\
 &&+ \frac{1}{(m_{P}^2-p^{\prime2})(m_{J/\psi}^2-p^2)} \int_{s^0_N}^\infty dt\frac{\rho^A_{PN^\prime}(p^{\prime 2},p^2,t)}{t-q^2}\nonumber\\
&& + \frac{1}{(m_{P}^2-p^{\prime2})(m_{N}^2-q^2)} \int_{s^0_{J/\psi}}^\infty dt\frac{\rho^A_{P\psi^\prime}(p^{\prime 2},t,q^2)}{t-p^2}  \nonumber\\
&& + \frac{1}{(m_{J/\psi}^2-p^{2})(m_{N}^2-q^2)} \int_{s^0_{P}}^\infty dt\frac{\rho^A_{P^{\prime}J/\psi}(t,p^2,q^2)+\rho^A_{P^{\prime}N}(t,p^2,q^2)}{t-p^{\prime2}}+\cdots \, ,\nonumber\\
\end{eqnarray}

\begin{eqnarray}\label{Hadron-PiB}
 \Pi_B(p^{\prime2},p^2,q^2)   & = & f_{J/\psi}m_{J/\psi}\lambda_P\lambda_N\frac{\left(m_P-m_N\right)g_V-g_T\frac{m_{J/\psi}^2}{m_P+m_N}}{\left(m_P^2-p^{\prime2}\right)\left(m_{J/\psi}^2-p^{2}\right)\left(m_{N}^2-q^{2}\right)}\nonumber\\
 &&+ \frac{1}{(m_{P}^2-p^{\prime2})(m_{J/\psi}^2-p^2)} \int_{s^0_N}^\infty dt\frac{\rho^B_{PN^\prime}(p^{\prime 2},p^2,t)}{t-q^2}\nonumber\\
&& + \frac{1}{(m_{P}^2-p^{\prime2})(m_{N}^2-q^2)} \int_{s^0_{J/\psi}}^\infty dt\frac{\rho^B_{P\psi^\prime}(p^{\prime 2},t,q^2)}{t-p^2}  \nonumber\\
&& + \frac{1}{(m_{J/\psi}^2-p^{2})(m_{N}^2-q^2)} \int_{s^0_{P}}^\infty dt\frac{\rho^B_{P^{\prime}J/\psi}(t,p^2,q^2)+\rho^B_{P^{\prime}N}(t,p^2,q^2)}{t-p^{\prime2}}+\cdots \, , \nonumber\\
\end{eqnarray}
where we introduce the formal functions $\rho^5_{PN^\prime}(p^{\prime 2},p^2,t)$, \, $\rho^5_{P\eta_c^\prime}(p^{\prime 2},t,q^2)$, $\rho^5_{P^{\prime}\eta_c}(t,p^2,q^2)$, \, $\rho^5_{P^{\prime}N}(t,p^2,q^2)$, \,$\overline{\rho}^5_{PN^\prime}(p^{\prime 2},p^2,t)$, \, $\overline{\rho}^5_{P\eta_c^\prime}(p^{\prime 2},t,q^2)$, $\overline{\rho}^5_{P^{\prime}\eta_c}(t,p^2,q^2)$, \, $\overline{\rho}^5_{P^{\prime}N}(t,p^2,q^2)$, \,$\rho^A_{PN^\prime}(p^{\prime 2},p^2,t)$, \,$\rho^A_{P\psi^\prime}(p^{\prime 2},t,q^2)$, \,$\rho^A_{P^{\prime}J/\psi}(t,p^2,q^2)$, \,$\rho^A_{P^{\prime}N}(t,p^2,q^2)$, \,$\rho^B_{PN^\prime}(p^{\prime 2},p^2,t)$, \,$\rho^B_{P\psi^\prime}(p^{\prime 2},t,q^2)$, \,$\rho^B_{P^{\prime}J/\psi}(t,p^2,q^2)$, and $\rho^B_{P^{\prime}N}(t,p^2,q^2)$ to parameterize the transitions between the ground states and the excited states. The $s^0_{\eta_c}$, $s^0_{J/\psi}$, $s^0_{N}$ and $s^0_P$ are the threshold parameters for the radial  excited states.

Now we smear the indexes $5$, $A$, $B$, et al,  and rewrite  (components of) the correlation functions  $\Pi_H(p^{\prime 2},p^2,q^2)$ at the hadron  side as
\begin{eqnarray}
\Pi_{H}(p^{\prime 2},p^2,q^2)&=&\int_{(m_{\eta_c}+m_N)^2}^{s_{P}^0}ds^\prime \int_{4m_c^2}^{s^0_{\eta_c}}ds \int_{0}^{s^0_{N}}du  \frac{\rho_H(s^\prime,s,u)}{(s^\prime-p^{\prime2})(s-p^2)(u-q^2)}\nonumber\\
&&+\int_{s^0_P}^{\infty}ds^\prime \int_{4m_c^2}^{s^0_{\eta_c}}ds \int_{0}^{s^0_{N}}du  \frac{\rho_H(s^\prime,s,u)}{(s^\prime-p^{\prime2})(s-p^2)(u-q^2)}+\cdots\, ,
\end{eqnarray}
 through dispersion relation,  and take $\eta_c=\eta_c$, $J/\psi$ for simplicity, where the $\rho_H(s^\prime,s,u)$   are the hadronic spectral densities.

We carry out the operator product expansion at the QCD side, and write  (components of) the correlation functions  $\Pi_{QCD}(p^{\prime 2},p^2,q^2)$  as
\begin{eqnarray}
\Pi_{QCD}(p^{\prime 2},p^2,q^2)&=&  \int_{4m_c^2}^{s^0_{\eta_c}}ds \int_{0}^{s^0_{N}}du  \frac{\rho_{QCD}(p^{\prime2},s,u)}{(s-p^2)(u-q^2)}+\cdots\, ,
\end{eqnarray}
through dispersion relation, where the $\rho_{QCD}(p^{\prime 2},s,u)$   are the QCD spectral densities, because the QCD spectral densities $\rho_{QCD}(s^\prime,s,u)$ do  not exist,
\begin{eqnarray}\label{QCD-s-prime}
\rho_{QCD}(s^\prime,s,u)&=&{\lim_{\epsilon\to 0}}\,\,\frac{ {\rm Im}_{s^\prime}\,\Pi_{QCD}(s^\prime+i\epsilon,s,u) }{\pi} =0\, ,
\end{eqnarray}
 we can  write the QCD spectral densities  $\rho_{QCD}(p^{\prime 2},s,u)$ as $\rho_{QCD}(s,u)$ for simplicity.

Now we match the hadron side  with the QCD side of the correlation functions,
and carry out the integral over $ds^\prime$  firstly to obtain the solid duality \cite{WangZhang-Solid},
\begin{eqnarray}
\int_{4m_c^2}^{s^0_{\eta_c}}ds \int_{0}^{s^0_{N}}du \frac{\rho_{QCD}(s,u)}{(s-p^2)(u-q^2)}&=&\int_{4m_c^2}^{s^0_{\eta_c}}ds \int_{0}^{s^0_{N}}du \frac{1}{(s-p^2)(u-q^2)}\left[ \int_{(\eta_c+m_N)^2}^{\infty} ds^\prime \frac{\rho_{H}(s^{\prime},s,u)}{s^\prime-p^{\prime2}}\right]\, . \nonumber\\
\end{eqnarray}

 It is impossible to carry out the  integral over $s^\prime$ explicitly due to the unknown functions  $\rho^5_{P^{\prime}\eta_c}(t,p^2,q^2)$, \, $\rho^5_{P^{\prime}N}(t,p^2,q^2)$,  \,$\overline{\rho}^5_{P^{\prime}\eta_c}(t,p^2,q^2)$, \, $\overline{\rho}^5_{P^{\prime}N}(t,p^2,q^2)$,  \,$\rho^A_{P^{\prime}J/\psi}(t,p^2,q^2)$, \,$\rho^A_{P^{\prime}N}(t,p^2,q^2)$,  \,$\rho^B_{P^{\prime}J/\psi}(t,p^2,q^2)$, and $\rho^B_{P^{\prime}N}(t,p^2,q^2)$. Now we introduce the parameters $C_5$, $\overline{C}_5$, $C_A$ and $C_B$ to parameterize the net effects,
\begin{eqnarray}
 C_5   & = &  \int_{s^0_{P}}^\infty dt\frac{\rho^5_{P^{\prime}\eta_c}(t,p^2,q^2)+\rho^5_{P^{\prime}N}(t,p^2,q^2)}{t-p^{\prime2}}\, , \nonumber\\
 \overline{C}_5   & = &  \int_{s^0_{P}}^\infty dt\frac{\overline{\rho}^5_{P^{\prime}\eta_c}(t,p^2,q^2)+\overline{\rho}^5_{P^{\prime}N}(t,p^2,q^2)}{t-p^{\prime2}}\, , \nonumber\\
 C_A   & = &  \int_{s^0_{P}}^\infty dt\frac{\rho^A_{P^{\prime}J/\psi}(t,p^2,q^2)+\rho^A_{P^{\prime}N}(t,p^2,q^2)}{t-p^{\prime2}} \, , \nonumber\\
 C_B   & = &  \int_{s^0_{P}}^\infty dt\frac{\rho^B_{P^{\prime}J/\psi}(t,p^2,q^2)+\rho^B_{P^{\prime}N}(t,p^2,q^2)}{t-p^{\prime2}}  \, .
\end{eqnarray}

In the following,  we write down  the quark-hadron duality explicitly,
 \begin{eqnarray}
  \int_{4m_c^2}^{s^0_{\eta_c}}ds \int_{0}^{s^0_{N}}du  \frac{\rho^5_{QCD}(s,u)}{(s-p^2)(u-q^2)}
  &=&\frac{f_{\eta_c}m_{\eta_c}^2\lambda_P\lambda_N}{2m_c}\frac{g_5}{\left(m_P^2-p^{\prime2}\right)\left(m_{\eta_c}^2-p^{2}\right)\left(m_{N}^2-q^{2}\right)}\nonumber\\
  &&+\frac{C_{5}}{(m_{\eta_c}^2-p^{2})(m_{N}^2-q^2)} \, ,
  \end{eqnarray}

  \begin{eqnarray}
  \int_{4m_c^2}^{s^0_{\eta_c}}ds \int_{0}^{s^0_{N}}du  \frac{\overline{\rho}^5_{QCD}(s,u)}{(s-p^2)(u-q^2)}
  &=&\frac{f_{\eta_c}m_{\eta_c}^2\lambda_P\lambda_N}{2m_c}\frac{\left(m_P+m_N\right)g_5}{\left(m_P^2-p^{\prime2}\right)\left(m_{\eta_c}^2-p^{2}\right)\left(m_{N}^2-q^{2}\right)}\nonumber\\
  &&+\frac{\overline{C}_{5}}{(m_{\eta_c}^2-p^{2})(m_{N}^2-q^2)} \, ,
  \end{eqnarray}

  \begin{eqnarray}
  \int_{4m_c^2}^{s^0_{J/\psi}}ds \int_{0}^{s^0_{N}}du  \frac{\rho^A_{QCD}(s,u)}{(s-p^2)(u-q^2)}
  &=&f_{J/\psi}m_{J/\psi}\lambda_P\lambda_N \frac{g_T-g_V}{\left(m_P^2-p^{\prime2}\right)\left(m_{J/\psi}^2-p^{2}\right)\left(m_{N}^2-q^{2}\right)}\nonumber\\
  &&+\frac{C_{A}}{(m_{J/\psi}^2-p^{2})(m_{N}^2-q^2)} \, ,
\end{eqnarray}

 \begin{eqnarray}
  \int_{4m_c^2}^{s^0_{J/\psi}}ds \int_{0}^{s^0_{N}}du  \frac{\rho^B_{QCD}(s,u)}{(s-p^2)(u-q^2)}
  &=&f_{J/\psi}m_{J/\psi}\lambda_P\lambda_N \frac{\left(m_P-m_N\right)g_V-g_T\frac{m_{J/\psi}^2}{m_P+m_N}}{\left(m_P^2-p^{\prime2}\right)\left(m_{J/\psi}^2-p^{2}\right)\left(m_{N}^2-q^{2}\right)}\nonumber\\
  &&+\frac{C_{B}}{(m_{J/\psi}^2-p^{2})(m_{N}^2-q^2)} \, .
\end{eqnarray}

 We set  $p^{\prime2}=p^2$  and perform  double Borel transform  with respect to the variables $P^2=-p^2$ and $Q^2=-q^2$, respectively  to obtain the  QCD sum rules,
\begin{eqnarray}\label{QCDSR-g5-Sgm}
&&\frac{f_{\eta_c}m_{\eta_c}^2\lambda_P\lambda_N}{2m_c} \frac{g_{5}}{m_{P}^2-m_{\eta_c}^2} \left[ \exp\left(-\frac{m_{\eta_c}^2}{T_1^2} \right)-\exp\left(-\frac{m_{P}^2}{T_1^2} \right)\right]\exp\left(-\frac{m_{N}^2}{T_2^2} \right) +\nonumber\\
&&C_{5} \exp\left(-\frac{m_{\eta_c}^2}{T_1^2} -\frac{m_{N}^2}{T_2^2} \right)=\int_{4m_c^2}^{s_{\eta_c}^0} ds \int_{0}^{s_N^0} du\, \rho^5_{QCD}(s,u)\exp\left(-\frac{s}{T_1^2} -\frac{u}{T_2^2} \right)\, ,
\end{eqnarray}

\begin{eqnarray}\label{QCDSR-g5-Gmu}
&&\frac{f_{\eta_c}m_{\eta_c}^2\lambda_P\lambda_N}{2m_c} \frac{\left(m_P+m_N \right)g_{5}}{m_{P}^2-m_{\eta_c}^2} \left[ \exp\left(-\frac{m_{\eta_c}^2}{T_1^2} \right)-\exp\left(-\frac{m_{P}^2}{T_1^2} \right)\right]\exp\left(-\frac{m_{N}^2}{T_2^2} \right) +\nonumber\\
&&\overline{C}_{5} \exp\left(-\frac{m_{\eta_c}^2}{T_1^2} -\frac{m_{N}^2}{T_2^2} \right)=\int_{4m_c^2}^{s_{\eta_c}^0} ds \int_{0}^{s_N^0} du\, \overline{\rho}^5_{QCD}(s,u)\exp\left(-\frac{s}{T_1^2} -\frac{u}{T_2^2} \right)\, ,
\end{eqnarray}

\begin{eqnarray}\label{QCDSR-gV-gT}
&&f_{J/\psi}m_{J/\psi}\lambda_P\lambda_N \frac{g_{V/T}}{m_{P}^2-m_{J/\psi}^2} \left[ \exp\left(-\frac{m_{J/\psi}^2}{T_1^2} \right)-\exp\left(-\frac{m_{P}^2}{T_1^2} \right)\right]\exp\left(-\frac{m_{N}^2}{T_2^2} \right) +\nonumber\\
&&C_{V/T} \exp\left(-\frac{m_{J/\psi}^2}{T_1^2} -\frac{m_{N}^2}{T_2^2} \right)=\int_{4m_c^2}^{s_{J/\psi}^0} ds \int_{0}^{s_N^0} du\, \rho^{V/T}_{QCD}(s,u)\exp\left(-\frac{s}{T_1^2} -\frac{u}{T_2^2} \right)\, ,
\end{eqnarray}

\begin{eqnarray}
C_V&=&\left[\frac{m_{J/\psi}^2}{m_P+m_N}C_A+C_B \right]\frac{m_P+m_N}{m_P^2-m_{J/\psi}^2-m_N^2}\, ,\nonumber\\
C_T&=&\left[\left(m_P-m_N\right)C_A+C_B \right]\frac{m_P+m_N}{m_P^2-m_{J/\psi}^2-m_N^2}\, ,\nonumber\\
\rho_{QCD}^V(s,u)&=&\left[\frac{m_{J/\psi}^2}{m_P+m_N}\rho_{QCD}^A(s,u)+\rho_{QCD}^B(s,u) \right]\frac{m_P+m_N}{m_P^2-m_{J/\psi}^2-m_N^2}\, ,\nonumber\\
\rho_{QCD}^T(s,u)&=&\left[\left(m_P-m_N\right)\rho_{QCD}^A(s,u)+\rho_{QCD}^B(s,u) \right]\frac{m_P+m_N}{m_P^2-m_{J/\psi}^2-m_N^2}\, ,
\end{eqnarray}

\begin{eqnarray}
\rho^5_{QCD}(s,u)&=&\frac{m_{c}}{4096\pi^6} \int_{x_{i}}^{x_{f}}dx\, u^2   -\frac{m_{c}^{3}}{36864\pi^4T_{1}^{4}} \langle\frac{\alpha_{s}GG}{\pi}\rangle  \int_{x_{i}}^{x_{f}}dx\, \frac{1}{x^3}\,u^2\delta\left(s-\widetilde{m}_c^2\right)
  \nonumber\\
&&+\frac{m_{c}}{12288\pi^4 T_{1}^{2}} \langle\frac{\alpha_{s}GG}{\pi}\rangle \int_{x_{i}}^{x_{f}}dx
 \frac{1-x}{x^2}\, u^2 \delta\left(s-\widetilde{m}_c^2\right)   \nonumber\\
 &&+\frac{m_{c}}{24576\pi^4T_{1}^{2}} \langle\frac{\alpha_{s}GG}{\pi}\rangle \int_{x_{i}}^{x_{f}}dx
 \frac{1}{x\left(1-x\right)} u^2\,\delta\left(s-\widetilde{m}_c^2\right)  \nonumber\\
 &&+ \frac{m_{c}}{2048\pi^4} \langle\frac{\alpha_{s}GG}{\pi}\rangle\int_{x_{i}}^{x_{f}}dx   -\frac{m_{c}}{2304\pi^4} \langle\frac{\alpha_{s}GG}{\pi}\rangle\int_{x_{i}}^{x_{f}}dx\,\frac{1}{x}u\,\delta\left(s-\widetilde{m}_c^2\right) \nonumber\\
&&+\frac{m_{c}\langle\bar{q}q\rangle \langle\bar{q}g_{s}\sigma Gq\rangle}{576\pi^2}\int_{x_{i}}^{x_{f}}dx\frac{1}{x(1-x)}\delta\left(s-\widetilde{m}_c^2\right)\delta(u)
   \nonumber\\
&&+\frac{m_{c}\langle\bar{q}q\rangle \langle\bar{q}g_{s}\sigma Gq\rangle}{576\pi^2 T_2^2}\int_{x_{i}}^{x_{f}}dx\, \delta(u)
   \nonumber\\
&&+\frac{m_{c}\langle\bar{q}g_{s}\sigma Gq\rangle^2}{2304\pi^2T_1^2}  \int_{x_{i}}^{x_{f}}dx
 \frac{1}{x\left(1-x\right)} \delta\left(s-\widetilde{m}_c^2\right)\delta(u)   \, ,
\end{eqnarray}

\begin{eqnarray}
\overline{\rho}^5_{QCD}(s,u)&=&  \frac{1}{2048\pi^6}\int_{x_{i}}^{x_{f}}dx\, x su^2  +\frac{m_{c}\langle\bar{q}q\rangle}{384\pi^4} \int_{x_{i}}^{x_{f}}dx\, u  +\frac{m_{c}\langle\bar{q}g_{s}\sigma Gq\rangle}{768\pi^4}  \int_{x_{i}}^{x_{f}}dx
 \frac{u}{x} \delta\left(s-\widetilde{m}_c^2\right)\nonumber\\
 &&+\frac{\langle\bar{q}q\rangle^2}{48\pi^2} \int_{x_{i}}^{x_{f}}dx\, x s\, \delta(u)   +\frac{m_{c}^{2}}{18432\pi^4T_1^2} \langle\frac{\alpha_{s}GG}{\pi}\rangle \int_{x_{i}}^{x_{f}}dx
\frac{u^2}{x^2} \left(2-\frac{s}{T_{1}^{2}}\right) \delta\left(s-\widetilde{m}_c^2\right) \nonumber\\
&&+\frac{1}{768\pi^4} \langle\frac{\alpha_{s}GG}{\pi}\rangle \int_{x_{i}}^{x_{f}}dx\, us\, \delta\left(s-\widetilde{m}_c^2\right) +\frac{1}{3072\pi^4}\langle\frac{\alpha_{s}GG}{\pi}\rangle \int_{x_{i}}^{x_{f}}dx\left(u+3xs\right)\nonumber\\
&&+\frac{1}{12288\pi^4}\langle\frac{\alpha_{s}GG}{\pi}\rangle \int_{x_{i}}^{x_{f}}dx\,(1+x)
\left(2us+u^2\right)\delta\left(s-\widetilde{m}_c^2\right) \nonumber\\
&&+\frac{1}{12288\pi^4T_1^2}\langle\frac{\alpha_{s}GG}{\pi}\rangle \int_{x_{i}}^{x_{f}}dx
\frac{2-x}{1-x} \,su^2\, \delta\left(s-\widetilde{m}_c^2\right)  \nonumber\\
&&-\frac{m_{c}^{3}}{3456\pi^2T_{1}^{4}}\langle\bar{q}q\rangle \langle\frac{\alpha_{s}GG}{\pi}\rangle
 \int_{x_{i}}^{x_{f}}dx \frac{u}{x^3}  \delta\left(s-\widetilde{m}_c^2\right)   \nonumber\\
&&+\frac{m_{c}\langle\bar{q}q\rangle}{1152\pi^2T_{1}^{2}} \langle\frac{\alpha_{s}GG}{\pi}\rangle
\int_{x_{i}}^{x_{f}}dx\frac{1-x}{x^2}   \,  u\,\delta\left(s-\widetilde{m}_c^2\right)   \nonumber\\
&&+ \frac{m_{c}\langle\bar{q}q\rangle}{2304\pi^2T_1^2} \langle\frac{\alpha_{s}GG}{\pi}\rangle\int_{x_{i}}^{x_{f}}dx
\frac{1}{x\left(1-x\right)}u\delta\left(s-\widetilde{m}_c^2\right) \nonumber\\
&&-\frac{m_c\langle\bar{q}q\rangle}{1152\pi^2} \langle\frac{\alpha_{s}GG}{\pi}\rangle \int_{x_{i}}^{x_{f}}dx\, \delta(u)\nonumber\\
&&+\frac{m_{c}^{2}\langle\bar{q}q\rangle^2}{432T_1^2} \langle\frac{\alpha_{s}GG}{\pi}\rangle
\int_{x_{i}}^{x_{f}}dx \frac{1}{x^2} \left(2-\frac{s}{T_{1}^{2}}\right)  \delta\left(s-\widetilde{m}_c^2\right)\delta(u)   \nonumber\\
&&+\frac{\langle\bar{q}q\rangle^2}{864T_{2}^{2}} \langle\frac{\alpha_{s}GG}{\pi}\rangle \int_{x_{i}}^{x_{f}}dx
\left[ 4s\delta\left(s-\widetilde{m}_c^2\right)+1  \right] \delta(u)\nonumber\\
&&+\frac{\langle\bar{q}q\rangle^2}{288T_{1}^{2}} \langle\frac{\alpha_{s}GG}{\pi}\rangle\int_{x_{i}}^{x_{f}}dx
\frac{2-x}{1-x}  \,s\,  \delta\left(s-\widetilde{m}_c^2\right) \delta(u)    \nonumber\\
&&-\frac{\langle\bar{q}q\rangle\langle\bar{q}g_{s}\sigma Gq\rangle}{96\pi^4T_2^2} \int_{x_{i}}^{x_{f}}dx\, xs \, \delta(u)+\frac{\langle\bar{q}q\rangle\langle\bar{q}g_{s}\sigma Gq\rangle}{288\pi^2} \int_{x_{i}}^{x_{f}}dx
\,s\,\delta\left(s-\widetilde{m}_c^2\right)\,\delta(u) \nonumber\\
&&+\frac{\langle\bar{q}q\rangle\langle\bar{q}g_{s}\sigma Gq\rangle}{576\pi^4} \int_{x_{i}}^{x_{f}}dx
\frac{1}{1-x}\,s\,\delta\left(s-\widetilde{m}_c^2\right) \,\delta(u)     \nonumber\\
&&+\frac{\langle\bar{q}q\rangle\langle\bar{q}g_{s}\sigma Gq\rangle}{384\pi^2T_2^2} \int_{x_{i}}^{x_{f}}dx
s\, \delta(u) +\frac{\langle\bar{q}g_{s}\sigma Gq\rangle^2}{2304\pi^2T_{2}^{2}}\int_{x_{i}}^{x_{f}}dx \,\delta(u)\nonumber\\
&&-\frac{\langle\bar{q}g_{s}\sigma Gq\rangle^2}{1728\pi^2T_2^2} \int_{x_{i}}^{x_{f}}dx\, s\, \delta\left(s-\widetilde{m}_c^2
\right) \,\delta(u)    -\frac{\langle\bar{q}g_{s}\sigma Gq\rangle^2}{13824\pi^2T_2^2}  \int_{x_{i}}^{x_{f}}dx \frac{1}{1-x}  \,s\, \delta\left(s-\widetilde{m}_c^2\right) \,\delta(u) \nonumber\\
&&+\frac{\langle\bar{q}g_{s}\sigma Gq\rangle^2}{13824\pi^2T_1^2} \int_{x_{i}}^{x_{f}}dx
\,s\,\delta\left(s-\widetilde{m}_c^2\right) \delta(u)-\frac{\langle\bar{q}g_{s}\sigma Gq\rangle^2}{13824\pi^2T_1^2} \int_{x_{i}}^{x_{f}}dx
\frac{1}{1-x}\,s\,  \delta\left(s-\widetilde{m}_c^2\right)  \delta(u)\nonumber\\
&&-\frac{\langle\bar{q}g_{s}\sigma Gq\rangle^2}{6912\pi^2} \int_{x_{i}}^{x_{f}}dx\,  \delta\left(s-\widetilde{m}_c^2\right)\delta(u) \, ,
\end{eqnarray}

\begin{eqnarray}
\rho^A_{QCD}(s,u)&=& \frac{m_{c}\langle\bar{q}q\rangle^2}{48\pi^2} \int_{x_{i}}^{x_{f}}dx\, \delta(u)  -\frac{5m_c\langle\bar{q}q\rangle\langle\bar{q}g_{s}\sigma Gq\rangle}{576\pi^2T_2^2} \int_{x_{i}}^{x_{f}}dx  \, \delta(u) \nonumber\\
 &&+\frac{m_c\langle\bar{q}q\rangle\langle\bar{q}g_{s}\sigma Gq\rangle}{288\pi^2} \int_{x_{i}}^{x_{f}}dx \frac{1}{x}\delta\left(s-\widetilde{m}_c^2\right)\, \delta(u) \nonumber\\
 &&+\frac{m_{c}}{9216\pi^4} \langle\frac{\alpha_{s}GG}{\pi}\rangle \int_{x_{i}}^{x_{f}}dx
\frac{1}{x(1-x)}u\, \delta\left(s-\widetilde{m}_c^2\right)    \nonumber\\
&&-\frac{m_{c}^3\langle\bar{q}q\rangle^2}{432T_{1}^{4}} \langle\frac{\alpha_{s}GG}{\pi}\rangle
\int_{x_{i}}^{x_{f}}dx\, \frac{1}{x^3}\delta\left(s-\widetilde{m}_c^2\right)\, \delta(u)   \nonumber\\
&&+\frac{m_{c}\langle\bar{q}q\rangle^2}{144T_1^2} \langle\frac{\alpha_{s}GG}{\pi}\rangle
\int_{x_{i}}^{x_{f}}dx\frac{1-x}{x^2}\delta\left(s-\widetilde{m}_c^2\right)\,\delta(u)   \nonumber\\
&&+\frac{m_{c} \langle\bar{q}q\rangle^2}{288T_2^2}  \langle\frac{\alpha_{s}GG}{\pi}\rangle\
 \int_{x_{i}}^{x_{f}}dx\left(\frac{1}{x}-\frac{2}{3}\right)
\delta\left(s-\widetilde{m}_c^2\right)\, \delta(u) \nonumber\\
&&+\frac{m_{c}\langle\bar{q}q\rangle^2}{864T_2^2} \langle\frac{\alpha_{s}GG}{\pi}\rangle\
 \int_{x_{i}}^{x_{f}}dx\frac{1-2x}{1-x}\delta\left(s-\widetilde{m}_c^2\right)\, \delta(u) \nonumber\\
&&-\frac{m_{c}\langle\bar{q}q\rangle^2}{864T_1^2} \langle\frac{\alpha_{s}GG}{\pi}\rangle\
 \int_{x_{i}}^{x_{f}}dx \frac{1}{x\left(1-x\right)} \delta\left(s-\widetilde{m}_c^2\right)\, \delta(u) \nonumber\\
  &&-\frac{m_{c}\langle\bar{q}g_{s}\sigma Gq\rangle^2}{4608\pi^2T_2^2} \int_{x_{i}}^{x_{f}}dx
\frac{1}{x(1-x)} \delta\left(s-\widetilde{m}_c^2\right) \, \delta(u)     \nonumber\\
   &&-\frac{m_{c}\langle\bar{q}g_{s}\sigma Gq\rangle^2}{13824\pi^2T_1^2} \int_{x_{i}}^{x_{f}}dx
 \frac{1}{x(1-x) } \delta\left(s-\widetilde{m}_c^2\right) \, \delta(u)   \, ,
\end{eqnarray}

\begin{eqnarray}
\rho^B_{QCD}(s,u)&=&-\frac{1}{4096\pi^6} \int_{x_{i}}^{x_{f}}dx\left[xs+x\left(1-x\right)\left(s-\widetilde{m}_c^2\right)\right]u^2 \nonumber\\
&&-\frac{\langle\bar{q}q\rangle^2}{24\pi^2} \int_{x_{i}}^{x_{f}}dx
\left[xs+x\left(1-x\right)\left(s-\widetilde{m}_c^2\right)\right]\,\delta(u) \nonumber\\
&&-\frac{m_{c}^2}{36864\pi^4T_{1}^{2}} \langle\frac{\alpha_{s}GG}{\pi}\rangle \int_{x_{i}}^{x_{f}}dx
\frac{1}{x^2}\left(2-\frac{s}{T_{1}^{2}}\right)\,u^2\,\delta\left(s-\widetilde{m}_c^2\right)  \nonumber\\
&&+\frac{m_{c}^2}{36864\pi^4T_{1}^{2}} \langle\frac{\alpha_{s}GG}{\pi}\rangle\
\int_{x_{i}}^{x_{f}}dx\frac{1-x}{x^2}u^2\delta\left(s-\widetilde{m}_c^2\right)  \nonumber\\
&&-\frac{1}{2304\pi^4} \langle\frac{\alpha_{s}GG}{\pi}\rangle \int_{x_{i}}^{x_{f}}dx
\left[\frac{(3-2x)us}{2(1-x)}\delta\left(s-\widetilde{m}_c^2\right)+u\right]           \nonumber\\
&&-\frac{1}{2048\pi^4} \langle\frac{\alpha_{s}GG}{\pi}\rangle \int_{x_{i}}^{x_{f}}dx
\left[\frac{4u}{9}+xs+x\left(1-x\right)\left(s-\widetilde{m}_c^2\right)\right]  \nonumber\\
&&+\frac{1}{73728\pi^4} \langle\frac{\alpha_{s}GG}{\pi}\rangle \int_{x_{i}}^{x_{f}}dx
\left(\frac{x}{1-x} \frac{s}{T_{1}^{2}}-1   \right) u^2\delta\left(s-\widetilde{m}_c^2\right)   \nonumber\\
&&+\frac{m_{c}^2\langle\bar{q}q\rangle^2}{216T_{1}^{2}}\langle\frac{\alpha_{s}GG}{\pi}\rangle\int_{x_{i}}^{x_{f}}dx
\left[-\frac{1}{x^2} \left(2-\frac{s}{T_{1}^{2}}\right)
+\frac{1-x}{x^2} \right]\delta\left(s-\widetilde{m}_c^2\right)\delta(u) \nonumber\\
&&- \frac{\langle\bar{q}q\rangle^2}{432T_2^2}  \langle\frac{\alpha_{s}GG}{\pi}\rangle
 \int_{x_{i}}^{x_{f}}dx\left[  \frac{(3-2x)s}{1-x}\delta\left(s-\widetilde{m}_c^2\right)+3   \right]\delta(u) \nonumber\\
&&+ \frac{\langle\bar{q}q\rangle^2}{432} \langle\frac{\alpha_{s}GG}{\pi}\rangle
 \int_{x_{i}}^{x_{f}}dx\left(\frac{x}{1-x}\frac{s}{T_{1}^{2}}-1\right)
\delta\left(s-\widetilde{m}_c^2\right)\delta(u)  \nonumber\\
&&-\frac{\langle\bar{q}q\rangle\langle\bar{q}g_{s}\sigma Gq\rangle}{576\pi^2}
 \int_{x_{i}}^{x_{f}}dx
\left[1+\frac{1+2x}{1-x}\,s\,\delta\left(s-\widetilde{m}_c^2\right)\right]\delta(u)  \nonumber\\
&&+\frac{\langle\bar{q}q\rangle\langle\bar{q}g_{s}\sigma Gq\rangle}{64\pi^2T_2^2} \int_{x_{i}}^{x_{f}}dx
\left[xs+x\left(1-x\right)\left(s-\widetilde{m}_c^2\right)\right]\,\delta(u)\nonumber\\
&&+\frac{ \langle\bar{q}g_{s}\sigma Gq\rangle^2}{2304\pi^2T_2^2}
 \int_{x_{i}}^{x_{f}}dx
\left[1+\frac{1+2x}{1-x}\,s\,\delta\left(s-\widetilde{m}_c^2\right)\right]\delta(u)  \, ,
\end{eqnarray}
where  $x_{f}=\frac{1+\sqrt{1-4m_{c}^{2}/s}}{2}$, $x_{i}=\frac{1-\sqrt{1-4m_{c}^{2}/s}}{2}$,   $\widetilde{m}_{c}^{2}=\frac{m_{c}^{2}}{x(1-x)}$, $\int_{x_{i}}^{x_{f}}dx\rightarrow\int_{0}^{1} dx$, when the $\delta$ function $\delta(s-\widetilde{m}_{c}^{2})$ appears.

In this article, we carry out the operator product expansion to the vacuum condensates  up to dimension-10, and  assume  vacuum saturation for the  higher dimension vacuum condensates. As the vacuum condensates are vacuum expectations of the quark-gluon operators, we take
the truncations $n\leq 10$ and $k\leq 1$ in a consistent way,
the operators of the orders $\mathcal{O}( \alpha_s^{k})$ with $k> 1$ are  neglected. Furthermore,  we set the two Borel parameters to be $T_1^2=T_2^2=T^2$ for simplicity, if we take the $T_1^2$ and $T_2^2$ as two independent parameters, it is difficult to obtain stable QCD sum rules.  In numerical calculations, we take the $C_5$, $\overline{C}_5$, $C_V$ and $C_T$ as free parameters and  choose the suitable values to obtain stable QCD sum rules.

\begin{figure}
 \centering
  \includegraphics[totalheight=4cm,width=5cm]{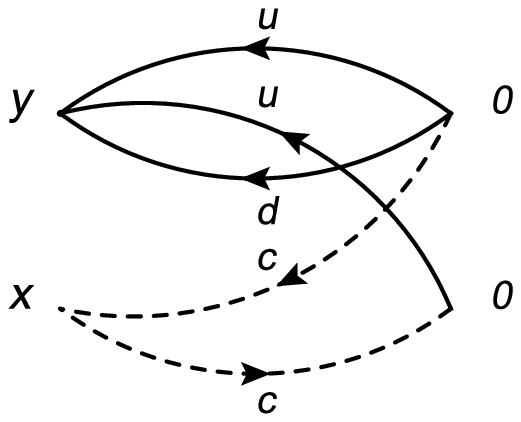}
  \hspace{1cm}
  \includegraphics[totalheight=4cm,width=5cm]{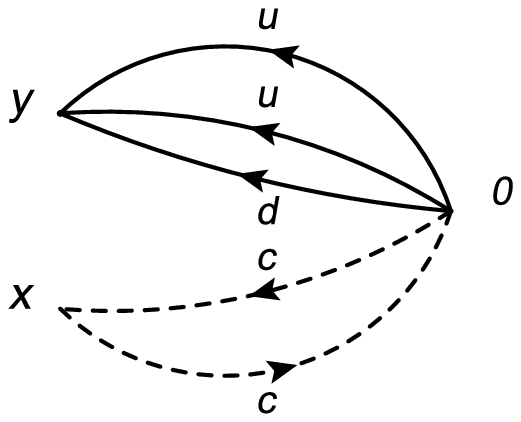}
 \caption{ The lowest order Feynman diagrams contributing  to decays of the  pentaquark molecular state. In the first diagram, we take into account the finite spatial separation between the  baryon and meson clusters. }\label{decay-per}
\end{figure}

\begin{figure}
 \centering
  \includegraphics[totalheight=4cm,width=5cm]{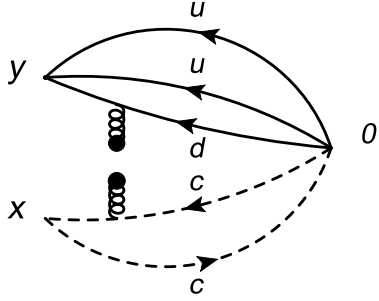}
 \caption{ The non-factorizable  Feynman diagrams contributing to the gluon condensate. Other
diagrams obtained by interchanging of the  light quark lines  or heavy quark lines are implied.}\label{decay-gg}
\end{figure}

In carrying out the operator product expansion for the correlation functions $\Pi_5(p,q)$ and $\Pi_{\mu}(p,q)$, if we take into account the finite spatial  separation between the clusters $\bar{c}(0)i\gamma_5 u(0)$ and $\varepsilon^{ijk}  u^T_i(0) C\gamma_\alpha d_j(0)\, \gamma^\alpha\gamma_5 c_{k}(0)$  in the current operator $J(0)$, the current $J(0)$ is modified to be
\begin{eqnarray}
J(0)&=& \bar{c}(\epsilon)i\gamma_5 u(\epsilon)\, \varepsilon^{ijk}  u^T_i(0) C\gamma_\alpha d_j(0)\, \gamma^\alpha\gamma_5 c_{k}(0)\, ,
\end{eqnarray}
by adding a small four-vector $\epsilon$, the Feynman diagrams for the decays to the charmonium states are non-factorizable, see the first Feynman diagram in Fig.\ref{decay-per}, where we split the point $0$ into two points to site the baryon and meson clusters respectively. In the limit $\epsilon \to 0$, the lowest order Feynman diagrams for the decays to the charmonium states are factorizable, see the
second Feynman diagram in Fig.\ref{decay-per}. In calculations, we observe that  there are both connected and disconnected Feynman diagrams  contributing to the decays,
the non-factorizable contributions begin at the order $\mathcal{O}(\sqrt{\alpha_s})$ due to the quark-gluon operators $\bar{q}g_s\sigma_{\alpha\beta}G^{\alpha\beta}q$, while at the order $\mathcal{O}(\alpha_s)$ of the quark-gluon operators, the non-factorizable contributions are of the forms $\langle \frac{\alpha_s}{\pi}GG\rangle$ and $\langle\bar{q}g_s\sigma Gq\rangle^2$. We absorb the strong coupling constant $g_s^2=4\pi\alpha_s$ into the vacuum condensates and count them as of the order $\mathcal{O}(\alpha_s^0)$. In Fig.\ref{decay-gg}, we draw the non-factorizable Feynman diagrams contributing to the gluon condensate as an example. Although the correlation functions $\Pi(p^{\prime2},p^2,p^2)$ can be written as
\begin{eqnarray}
\Pi(p^{\prime2},p^2,p^2)&=&\int_{4m_c^2}^{\infty}ds \int_{0}^{\infty}du \frac{\rho_{QCD}(s,u)}{(s-p^2)(u-q^2)}+{\rm nonsingular\,\, terms\,\, of}\,\, p^{\prime2} \, ,
\end{eqnarray}
at the QCD side, there are both factorizable and non-factorizable contributions.

In previous section, we have proved  that the current operator $J(x)$ couples potentially to the $\Sigma_c\bar{D}$ molecular state, which receives  both factorizable and non-factorizable contributions,  while the couplings to the baryon-meson scattering states  can be neglected. From Eqs.\eqref{Hadron-Pi5}-\eqref{Hadron-PiB}, we can see that there is a pole term $\frac{1}{m_P^2-p^{\prime2}}$ at the hadron side, which should have origins  at the QCD side, while at the QCD side, there is no singular term with respect to the variable $p^{\prime2}$, see Eq.\eqref{QCD-s-prime}. It does not mean that there is no contribution from the $P_c(4312)$ or the current-molecule coupling $\lambda_P$ is zero, it just means that the $P_c(4312)$ may be  not on the mass-shell. In fact, we set $p^{\prime2}=p^2$ to obtain the QCD sum rules,
the terms $\frac{1}{m_P^2-p^{2}}$ and $\frac{1}{m_{J/\psi}^2-p^{2}}$ at the hadron side cannot be singular simultaneously. The reasonable explanation is that the
current operator $J(0)$ in the three-point correlation functions $\Pi_5(p,q)$ and $\Pi_{\mu}(p,q)$ couples  potentially to the $\Sigma_c\bar{D}$ molecular state or the $P_c(4312)$, however, the $P_c(4312)$ may be not on the mass-shell, which facilitates  the trick of setting $p^{\prime2}=p^2$.

\section{Numerical results and discussions}
At the hadron side, we take the hadronic parameters  as
$m_{J/\psi}=3.0969\,\rm{GeV}$, $m_{N}=0.93827\,\rm{GeV}$,
$m_{\eta_c}=2.9839\,\rm{GeV}$,
$\sqrt{s^0_{J/\psi}}=3.6\,\rm{GeV}$, $\sqrt{s^0_{\eta_c}}=3.5\,\rm{GeV}$, $\sqrt{s^0_{N}}=1.3\,\rm{GeV}$
 \cite{PDG}, $m_{P}=4.3119\,\rm{GeV}$ \cite{LHCb-Pc4312},
$f_{J/\psi}=0.418 \,\rm{GeV}$, $f_{\eta_c}=0.387 \,\rm{GeV}$  \cite{Becirevic},
$\lambda_N=0.032\,\rm{GeV}^3$ \cite{SB-Ioffe}, $\lambda_P=1.95\times 10^{-3}\,\rm{GeV}^6$  \cite{WangPc4450-molecule}.

At the QCD side, we take  the standard values of the vacuum condensates $\langle
\bar{q}q \rangle=-(0.24\pm 0.01\, \rm{GeV})^3$,   $\langle
\bar{q}g_s\sigma G q \rangle=m_0^2\langle \bar{q}q \rangle$,
$m_0^2=(0.8 \pm 0.1)\,\rm{GeV}^2$,   $\langle \frac{\alpha_s
GG}{\pi}\rangle=(0.33\,\rm{GeV})^4 $    at the energy scale  $\mu=1\, \rm{GeV}$
\cite{PRT85,SVZ79,ColangeloReview}, and choose the $\overline{MS}$ mass  $m_{c}(m_c)=(1.275\pm0.025)\,\rm{GeV}$
 from the Particle Data Group \cite{PDG}.
Moreover, we take into account the energy-scale dependence of  the  parameters,
\begin{eqnarray}
\langle\bar{q}q \rangle(\mu)&=&\langle\bar{q}q \rangle({\rm 1 GeV})\left[\frac{\alpha_{s}({\rm 1 GeV})}{\alpha_{s}(\mu)}\right]^{\frac{12}{25}}\, , \nonumber\\
 \langle\bar{q}g_s \sigma Gq \rangle(\mu)&=&\langle\bar{q}g_s \sigma Gq \rangle({\rm 1 GeV})\left[\frac{\alpha_{s}({\rm 1 GeV})}{\alpha_{s}(\mu)}\right]^{\frac{2}{25}}\, , \nonumber\\
m_c(\mu)&=&m_c(m_c)\left[\frac{\alpha_{s}(\mu)}{\alpha_{s}(m_c)}\right]^{\frac{12}{25}} \, ,\nonumber\\
\alpha_s(\mu)&=&\frac{1}{b_0t}\left[1-\frac{b_1}{b_0^2}\frac{\log t}{t} +\frac{b_1^2(\log^2{t}-\log{t}-1)+b_0b_2}{b_0^4t^2}\right]\, ,
\end{eqnarray}
   where $t=\log \frac{\mu^2}{\Lambda^2}$, $b_0=\frac{33-2n_f}{12\pi}$, $b_1=\frac{153-19n_f}{24\pi^2}$,
   $b_2=\frac{2857-\frac{5033}{9}n_f+\frac{325}{27}n_f^2}{128\pi^3}$,
   $\Lambda=210\,\rm{MeV}$, $292\,\rm{MeV}$  and  $332\,\rm{MeV}$ for the flavors
   $n_f=5$, $4$ and $3$, respectively  \cite{PDG,Narison-mix}, and evolve all the parameters to the ideal   energy scale   $\mu$  with $n_f=4$ to extract the hadronic coupling constants $g_5$, $g_V$ and $g_T$.

  In the QCD sum rules for the mass of the $\bar{D}\Sigma_c$ pentaquark  molecular state with the spin-parity $J^P={\frac{1}{2}}^-$ or the $P_c(4312)$, the ideal  energy scale of the QCD spectral density is $\mu=2.2\,\rm{GeV}$ \cite{WangPc4450-molecule}, which is determined by the energy scale formula $\mu=\sqrt{M^2_{X/Y/Z/P}-(2{\mathbb{M}}_c)^2}$ with the effective $c$-quark mass ${\mathbb{M}}_c=1.85\,\rm{GeV}$ \cite{Wang-CPC-Mc}. The energy scale $\mu=2.2\,\rm{GeV}$ is tool large for the $N$, $\eta_c$ and $J/\psi$.
     In this article, we take the energy scales of the QCD spectral densities to be $\mu=\frac{m_{\eta_c}}{2}=1.5\,\rm{GeV}$, which is acceptable for the
     charmonium states \cite{WangHuangTao-3900}.

We choose the values of the free parameters  as
$C_{5}=1.18\times 10^{-6}\,\rm{GeV}^9 $, $\overline{C}_{5}=1.94\times 10^{-5}\,\rm{GeV}^{10} $,
$C_{V}=-1.77\times 10^{-5}\,\rm{GeV}^9  $,
$C_{T}=-1.67\times 10^{-5}\,\rm{GeV}^{9} $
   to obtain flat platforms in the Borel windows $T^2 =(3.1-4.1)\,\rm{GeV^2}$, $(3.3-4.3)\,\rm{GeV^2}$, $(4.0-5.0)\,\rm{GeV^2}$ and $(3.9-4.9)\,\rm{GeV^2}$ for the hadronic coupling constants $g_5$, $g_V$ and $g_T$, respectively. We fit the free parameters $C_5$,  $\overline{C}_5$, $C_V$ and $C_T$ to obtain the same intervals of flat  platforms $T_{max}^2-T^2_{min}=1.0\,\rm{GeV}^2$, where the $T^2_{max}$ and $T^2_{min}$ denote the maximum and minimum of the Borel parameters, respectively.

\begin{figure}
 \centering
  \includegraphics[totalheight=5cm,width=7cm]{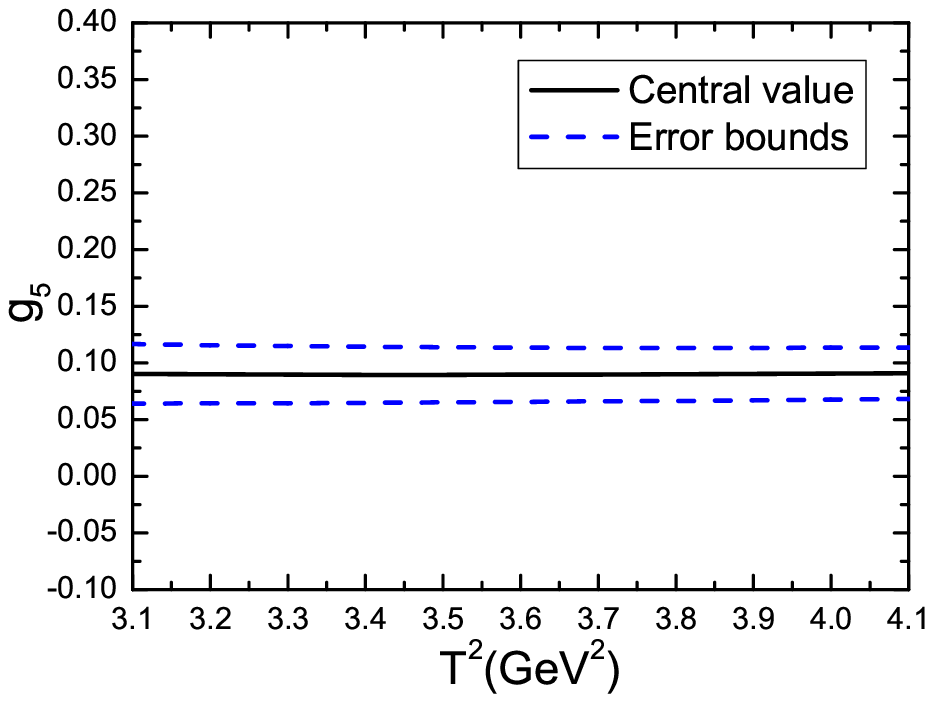}
  \includegraphics[totalheight=5cm,width=7cm]{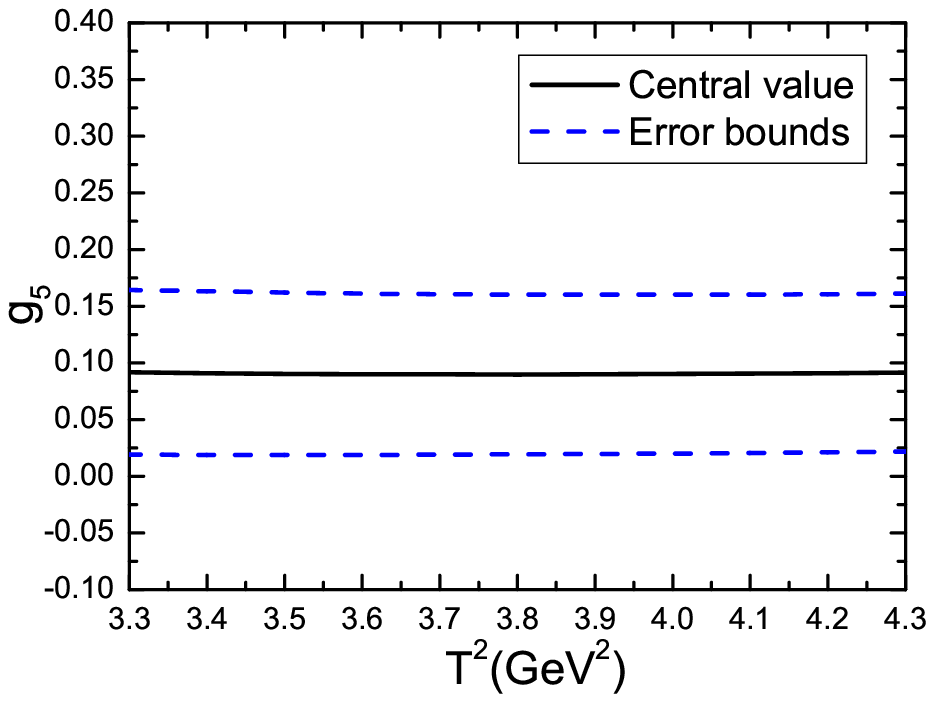}
   \includegraphics[totalheight=5cm,width=7cm]{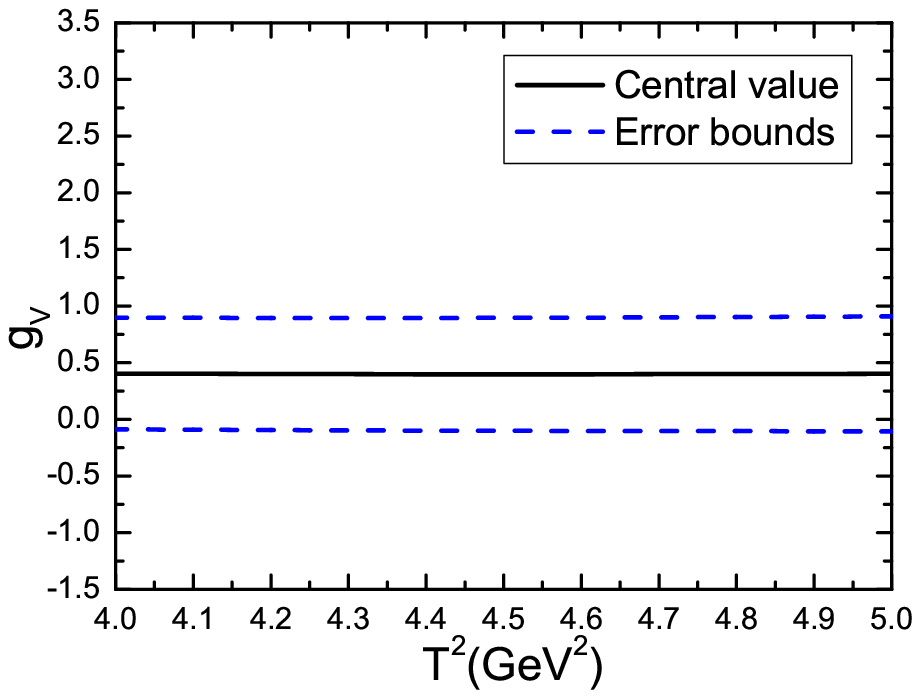}
   \includegraphics[totalheight=5cm,width=7cm]{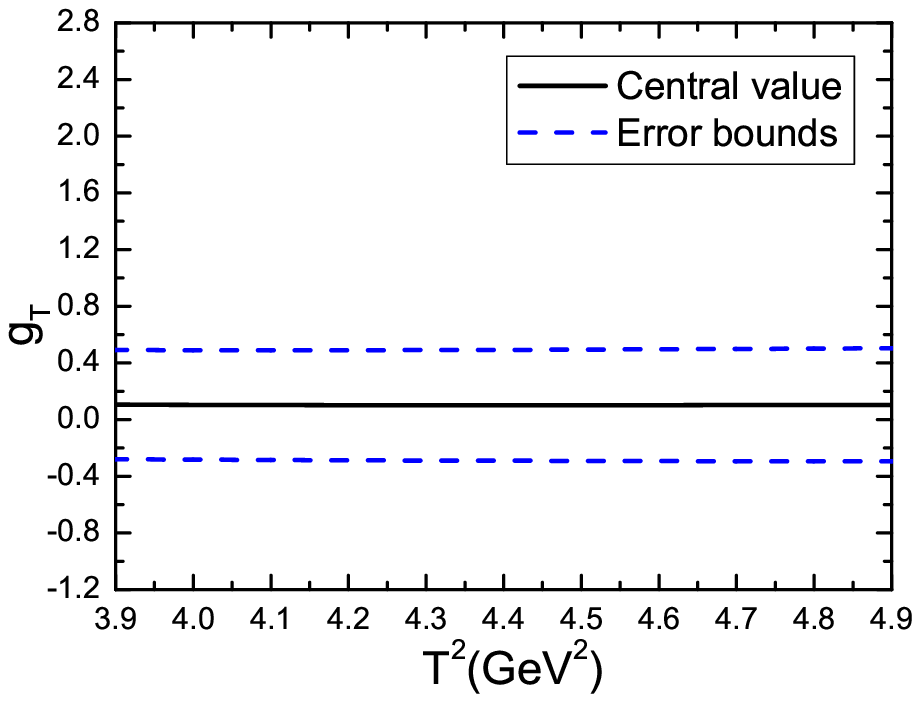}
     \caption{The hadronic coupling constants $g_5$, $g_V$ and $g_T$ with variations of the  Borel  parameters  $T^2$, the values of the $g_5$ in the first diagram and second diagram come from the QCD sum rules in Eq.\eqref{QCDSR-g5-Sgm} and Eq.\eqref{QCDSR-g5-Gmu}, respectively.}\label{Borel-G}
\end{figure}

We take into account  the uncertainties  of the input   parameters,
and obtain the values of  the hadronic coupling constants $g_5$, $g_V$ and $g_T$, which are shown in Fig.\ref{Borel-G},
\begin{eqnarray}
g_5&=&0.09\pm0.03\,\,\,\,{\rm from} \,\,\,\,{\rm Eq.}\eqref{QCDSR-g5-Sgm}\, ,\nonumber\\
g_5&=&0.09\pm0.07\,\,\,\,{\rm from}\,\,\,\,{\rm Eq.}\eqref{QCDSR-g5-Gmu}\, ,\nonumber\\
g_V&=&0.40\pm0.50\, ,\nonumber\\
g_T&=&0.10\pm0.40\, ,
\end{eqnarray}
where we have redefined  the hadronic coupling constants $g_V/g_T$ in Eq.\eqref{Coupling-gVgT} with a simple replacement $g_V/g_T \to -g_V/g_T$, as the central values of the $g_V/g_T$ are negative from the QCD sum rules in Eq.\eqref{QCDSR-gV-gT}.

Now it is straightforward to calculate the partial decay widths of the decays $P_c(4312)\to \eta_c N$, $J/\psi N$,
\begin{eqnarray}
\Gamma\left(P_c(4312)\to \eta_c N\right)&=& \frac{p(m_P,m_{\eta_c},m_N)}{16\pi m_P^2} |T|^2\nonumber\\
&=&31.488 g_5^2 \,\,{\rm{MeV}}\nonumber\\
&=&0.255^{+0.198}_{-0.142}\,{\rm{MeV}}\,\,\,\,{\rm from} \,\,\,\,{\rm Eq.}\eqref{QCDSR-g5-Sgm}\, ,\nonumber\\
&=&0.255^{+0.551}_{-0.242}\,{\rm{MeV}}\,\,\,\,{\rm from}\,\,\,\,{\rm Eq.}\eqref{QCDSR-g5-Gmu}\, ,
\end{eqnarray}
where
\begin{eqnarray}
T&=&\bar{u}(q)i g_5 u(p^\prime)\, ,
\end{eqnarray}
and $p(a,b,c)=\frac{\sqrt{[a^2-(b+c)^2][a^2-(b-c)^2]}}{2a}$,

\begin{eqnarray}
\Gamma\left(P_c(4312)\to J/\psi N\right)&=& \frac{p(m_P,m_{J/\psi},m_N)}{16\pi m_P^2} |T|^2\nonumber\\
&=&29.699 g_T^2 - 97.554g_V g_T + 80.633g_V^2\,\,{\rm{MeV}}\nonumber\\
&=&9.296^{+19.542}_{-9.296}\,\,{\rm{MeV}}\, ,
\end{eqnarray}
where
\begin{eqnarray}
T&=&\varepsilon_\alpha^*\bar{u}(q)\left( g_V\gamma^\alpha-i\frac{g_T}{m_P+m_N}\sigma^{\alpha\beta}p_\beta\right)\gamma_5 u(p^\prime)\, .
\end{eqnarray}

The partial decay width $\Gamma\left(P_c(4312)\to \eta_c N\right)=0.255\,{\rm{MeV}}$ is vary small, the total width $\Gamma_{P_c(4312)}$ can be saturated with the strong decay  $P_c(4312)\to J/\psi N$. The predicted width  $\Gamma\left(P_c(4312)\to J/\psi N\right)=9.296^{+19.542}_{-9.296}\,\,{\rm{MeV}}$ is compatible with the experimental data $\Gamma_{P_c(4312)} = 9.8\pm2.7^{+ 3.7}_{- 4.5} \mbox{ MeV}$ from the LHCb collaboration \cite{LHCb-Pc4312}. The present calculations support assigning the
$P_c(4312)$ to be the $\bar{D}\Sigma_c$ pentaquark molecular state with the spin-parity  $J^P={\frac{1}{2}}^-$. We can search for the $P_c(4312)$ in the $\eta_cN$ mass spectrum, and measure the branching fraction
${\rm Br}\left(P_c(4312)\to \eta_c N\right)$, which maybe shed light on the nature of the $P_c(4312)$ and test  the predictions of the QCD sum rules.

The thresholds of the $\bar{D}\Lambda_c$ and $\bar{D}^*\Lambda_c$ are $4.15\,\rm{GeV}$ and $4.29\,\rm{GeV}$, respectively, the decays to the final states $\bar{D}\Lambda_c$ and $\bar{D}^*\Lambda_c$ are kinematically allowed. At the quark level, the decays  of the $\bar{D}\Sigma_c$ pentaquark molecular state to the $\bar{D}\Lambda_c$ and $\bar{D}^*\Lambda_c$ states  take place through dissolving of the $\Sigma$-type diquark states to form the $\Lambda$-type diquark states by emitting an isospin $I=1$ quark-antiquark pair. At the hadron level, the decay $P_c(4312)\to\bar{D}^*\Lambda_c$ can take place through process     $P_c(4312)\to\bar{D}\Sigma_c \to\bar{D}\Lambda_c\pi \to \bar{D}^*\Lambda_c$  with the subprocesses  $\Sigma_c\to \pi\Lambda_c$ and $\bar{D}\pi\to\bar{D}^*$, the partial decay width $\Gamma(P_c(4312)\to \bar{D}^*\Lambda_c)$ may be as large as $10.7\,\rm{MeV}$ \cite{ZouBS-2019}. Direct calculations of those partial decay widths with the QCD sum rules are necessary to make a definite conclusion, this is our next work.

\section{Conclusion}
In this article, we tentatively assign the $P_c(4312)$ to be the $\bar{D}\Sigma_c$ pentaquark molecular state with the spin-parity $J^P={\frac{1}{2}}^-$, and discuss the
 factorizable  and non-factorizable contributions in the two-point QCD sum rules for  the $\bar{D}\Sigma_c$  molecular state in details to prove
 the reliability of the single pole approximation in the hadronic spectral density. We study its two-body strong decays with the QCD sum rules by carrying out the operator product expansion up to the vacuum condensates of dimension $10$.
In calculations, special attentions are paid to match the hadron side with the QCD side of the correlation functions  to obtain  solid duality.  We obtain the  partial decay widths $\Gamma\left(P_c(4312)\to \eta_c p\right)=0.255\,{\rm{MeV}}$ and $\Gamma\left(P_c(4312)\to J/\psi p\right)=9.296^{+19.542}_{-9.296}\,\,{\rm{MeV}}$,  which
are compatible with the experimental data $\Gamma_{P_c(4312)} = 9.8\pm2.7^{+ 3.7}_{- 4.5} \mbox{ MeV}$ from the LHCb collaboration. The present calculations support assigning the
$P_c(4312)$ to be the $\bar{D}\Sigma_c$ pentaquark molecular state with the spin-parity $J^P={\frac{1}{2}}^-$. We can search for the decay  $P_c(4312)\to \eta_c p$  to diagnose the nature of the $P_c(4312)$.

\section*{Acknowledgements}
This  work is supported by National Natural Science Foundation, Grant Number  11775079.

\end{document}